\documentclass[fleqn,usenatbib]{mnras}

\usepackage{newtxtext,newtxmath}

\usepackage[T1]{fontenc}

\DeclareRobustCommand{\VAN}[3]{#2}
\let\VANthebibliography\thebibliography
\def\thebibliography{\DeclareRobustCommand{\VAN}[3]{##3}\VANthebibliography}

\usepackage{graphicx}	
\usepackage{amsmath}	
\usepackage{flexisym}	
\usepackage{gensymb}
\usepackage{subfig}
\usepackage{newtxtext,newtxmath}

\newcommand{\psr}{PSR J1555--2908}

\title[The mass of \psr]{Measuring the mass of the black widow PSR J1555-2908}

\author[Kennedy et al.]{M. R. Kennedy$^{1,2}$\thanks{E-mail: markkennedy@ucc.ie}, R. P. Breton$^1$, C. J. Clark$^{1,3,4}$, D. Mata-Sanchez$^{1,5,6}$, G. Voisin$^{7,1}$, V. S. Dhillon$^{8,5}$,\newauthor J. P. Halpern$^{9}$,  T. R. Marsh$^{10}$, L. Nieder$^{3,4}$,  P. S. Ray$^{11}$, M. H. van Kerkwijk$^{12}$
\\
$^{1}$Jodrell Bank Centre for Astrophysics, Department of Physics and Astronomy, The University of Manchester, M19 9PL, UK\\
$^{2}$Department of Physics, University College Cork, Cork, Ireland\\
$^{3}$ Max-Planck-Institut f\"{u}r Gravitationsphysik (Albert-Einstein-Institut), Callinstra{\ss}e 38, 30167 Hannover, Germany\\
$^{4}$ Leibniz Universit\"{a}t Hannover, 30167 Hannover, Germany\\
$^{5}$Instituto de Astrof\'{i}sica de Canarias, E-38205 La Laguna, Tenerife, Spain \\
$^{6}$Departamento de Astrof\'{i}sica, Universidad de La Laguna, E-38206 La Laguna, Tenerife, Spain\\
$^{7}$LUTH, Observatoire de Paris, PSL Research University, CNRS, 5 Place Jules Janssen, 92190 Meudon, France\\
$^{8}$Department of Physics and Astronomy, University of Sheffield, Sheffield S3 7RH, UK \\
$^{9}$Department of Astronomy, Columbia University, 550 West 120th Street, New York, NY 10027, USA\\
$^{10}$Department of Physics, University of Warwick, Coventry CV4 7AL, UK\\
$^{11}$Space Science Division, U.S. Naval Research Laboratory, Washington, DC 20375, USA\\
$^{12}$Department of Astronomy and Astrophysics, University of Toronto, 50 George Street, Toronto, Ontario, M5S 3H4, Canada
}

\date{Accepted 2022 February 7. Received 2022 February 7; in original form 2021 November 8
}

\pubyear{2021}

\begin{document}
\label{firstpage}
\pagerange{\pageref{firstpage}--\pageref{lastpage}}
\maketitle
\begin{abstract}
Accurate measurements of the masses of neutron stars are necessary to test binary evolution models, and to constrain the neutron star equation of state. In pulsar binaries with no measurable post-Keplerian parameters, this requires an accurate estimate of the binary system's inclination and the radial velocity of the companion star by other means than pulsar timing. In this paper, we present the results of a new method for measuring this radial velocity using the binary synthesis code \textsc{Icarus}. This method relies on constructing a model spectrum of a tidally distorted, irradiated star as viewed for a given binary configuration. This method is applied to optical spectra of the newly discovered black widow \psr. By modelling the optical spectroscopy alongside optical photometry, we find that the radial velocity of the companion star is $397\pm4$ km s$^{-1}$ (errors quoted at 95\% confidence interval), as well as a binary inclination of $>75$\degree. Combined with $\gamma$-ray pulsation timing information, this gives a neutron star mass of 1.67$^{+0.15}_{-0.09}$ M$_\odot$ and a companion mass of 0.060$^{+0.005}_{-0.003}$ M$_\odot$, placing \psr\ at the observed upper limit of what is considered a black widow system. 
\end{abstract}

\begin{keywords}
stars: neutron -- binaries: close -- pulsars: individual: PSR J1555-2908 -- techniques: spectroscopic
\end{keywords}



\section{Introduction}
Since the discovery of PSR B1957+20 \citep{1988Natur.333..237F}, the so called ``black widow'' systems have become increasingly important in understanding neutron stars. These systems have rapidly rotating neutron star primaries (millisecond pulsars; MSPs) which are heating and ablating their nearby, extremely low mass companions ($M_{\rm Comp} <0.1$ M$_{\odot}$). They are mainly differentiated from their ``redback'' cousins by their companion mass, as redbacks have much heavier companions with ($M_{\rm Comp} >0.1 {\rm M}_{\odot}$, \citealt{2013ApJ...775...27C}). The underlying question of whether two distinct populations exist, or if redbacks evolve into black widows is still under intense debate \citep{2020MNRAS.495.3656G}.

The emergence of black widows as important systems in understanding neutron star physics was been heralded by two discoveries. The first is the measurement of the mass of the neutron star in PSR B1957+20 of $M_{\rm NS} = 2.40\pm0.12$ M$_{\odot}$ (\citealt{2011ApJ...728...95V}; assuming a binary inclination of 65\degree as measured by \citealt{2007MNRAS.379.1117R}; 1$\sigma$ error), making this neutron star one of the most massive known to date. The second is the discovery of the black widow PSR J0952--0607, whose 1.4 ms spin period is the second fastest of any known pulsar \citep{2017ApJ...846L..20B}. These two discoveries suggest that black widows are excellent candidates for hosting neutron stars with extreme properties. Most interestingly, black widows and redbacks appear to have a higher median neutron star mass than other systems containing NSs (\citealt{Strader2019}; \citealt{Linares2019}), making them ideal candidates for hunting for heavy neutron stars which may help to constrain the equation of state of nuclear matter.

Measuring the component masses in a binary requires combining the system's mass ratio,
\begin{equation}
    q=\frac{M_{1}}{M_{2}}=\frac{K_{2}}{K_{1}},
\end{equation}
where $M_{1}$ and $M_{2}$ are the component masses and $K_{1}$ and $K_{2}$ are the radial velocity amplitudes for star 1 and star 2 respectively, with the mass function
\begin{equation}
    \frac{M_1\sin^3{i}}{(1+1/q)^2}=\frac{P_{\rm orb}K^3_2}{2\pi G}
\end{equation}
where $i$ is the binary inclination, $P_{\rm orb}$ is the orbital period, and G is Newton's gravitational constant. At face value, this seems straightforward. The pulsar's radial velocity amplitude can be measured from the delay in the arrival time of pulses across the binary orbit. The companion's radial velocity amplitude can be estimated by measuring the motion of absorption lines from the companion in optical and infra-red spectra of the binary. Finally, the orbital inclination can be obtained through modelling of the optical light curve.

However, inferring accurate masses in these binary systems is becoming an increasingly arduous task. In recent years, numerous light curves of black widows and redbacks have been identified with asymmetries that cannot be explained solely by uniform illumination and heating of the companion stars in these systems. Several models have been invoked to explain these asymmetries including: stellar spots on the surface of the companion (\citealt{2016ApJ...833L..12V}; \citealt{2020MNRAS.tmp.3423C}); additional optical light due to the effects from an intra-binary shock \citep{2016ApJ...828....7R}; channeling of material onto the magnetic poles of the companion \citep{2017ApJ...845...42S}; heat redistribution across the surface of the companion through convection and diffusion (\citealt{2020MNRAS.tmp.2692V}; \citealt{2020ApJ...892..101K}). Which of these models is appropriate for any given system is unclear, but results from these studies show that the recovered binary parameters, particularly the system's inclination, can be heavily biased depending on which effects are included when modelling the light curve. Coupled with the difficulties encountered when disentangling the projected radial velocity amplitude of a star's centre of mass from the radial velocity of its centre of light in the presence of strong irradiation gives constituent masses with large systematic uncertainties. To correct these, \citealt{2011ApJ...728...95V} relied on estimates of the effect of the distribution of light, while \citealt{Linares2018} took an empirical approach, bracketing the true velocities using measurements from lines selected from their temperature sensitivity to arise preferably on the inner or outer hemispheres of the companion star.  Our work follows on both approaches, matching observed spectra with model spectra integrated over the visible face of the companion.

One of the systems discovered in the hunt for new black widows, and the subject of this paper, is PSR J1555--2908. This system was originally identified as a pulsar candidate in a search for steep spectrum radio continuum sources coincident with unassociated $\gamma$--ray sources from a preliminary 7-year Fermi Large Area Telescope source catalogue \citep{2018MNRAS.475..942F}. A follow-up radio search with the Green Bank Telescope revealed it to be a millisecond pulsar with a low-mass companion and the detection of an optical counterpart confirmed its black widow nature (Ray et al. 2021, submitted.). The detection of gamma-ray pulsations using the radio timing ephemeris provided a 12-year timing solution, in which the pulsar's spin-down rate could be measured, revealing it to be one of the most energetic millisecond pulsars in  the Galaxy. Combined with its orbital period of 5.6 hr, \psr\ provides a unique test bed for investigating the effects of extreme irradiation of the companion star.

Here, we present optical photometry and spectroscopy of this new spider system. Section~\ref{sec:Obs} discusses the observations, while Section~\ref{sec:modelling} focuses on modelling of the optical data, using a method for modelling the optical spectrum in a self-consistent way with the photometry for the first time. Finally, Section~\ref{sec:Disc} discusses the implications of the binary parameters and Section~\ref{sec:Conc} summarises the work.

Finally, through out the paper, errors are quoted the 95\% confidence level, unless explicitly stated. This is largely due to the fact that the posterior distributions on the samples for many parameters when performing error analysis were non-Gaussian. As such, the commonly used 1$\sigma$ errors are not appropriate in accurately describing the full confidence range for these parameters.

\section{Observations}\label{sec:Obs}

\subsection{Optical photometry}
\psr\ was observed over several nights in 2018 and 2019 using ULTRACAM \citep{dhillon2007} mounted on the European Southern Observatory's 3.5m New Technology Telescope (NTT) as detailed in Table~\ref{tab:obs_details}. ULTRACAM provides three simultaneous bands of optical photometry with a readout time $\sim$ 24 ms. The data were acquired with Super-SDSS $u_{\rm s}$, $g_{\rm s}$, and $i_{\rm s}$ filters, which are filters that cover the same wavelength range as the traditional SDSS \textit{u\textprime}, \textit{g\textprime}, and \textit{i\textprime}\ filters \citep{2010AJ....139.1628D}, but with a higher throughput \citep{HCAM}. The exposure times on each night of data were the same, with one frame read out every 10 s in both $g_{\rm s}$ and $i_{\rm s}$ channels, and one frame read out every 30 s in $u_{\rm s}$ to increase the signal-to-noise ratio in this channel.

The data were reduced using the ULTRACAM pipeline \citep{dhillon2007}, with night-to-night variations in the zeropoint and transparency corrected using ``ensemble photometry'' \citep{1992PASP..104..435H} based on 5 nearby stars present during each observation. 

\begin{table}
    \centering
    \caption{Details of the optical photometry and spectroscopy taken of \psr\ using ULTRACAM on the NTT and X-SHOOTER on Unit 2 of the VLT. Orbital phase has been calculated using the timing parameters given in Table~\ref{tab:results}. There are small gaps in both of the X-SHOOTER runs where additional wavelength calibration spectra were obtained as detailed in the text.}
    \label{tab:obs_details}
    \begin{tabular}{c|c c}
    \hline
    Start Time (UTC) & Duration (hr) & Phase Coverage\\
    \hline
    \multicolumn{3}{c|}{ULTRACAM+NTT (Photometry)}\\
    \hline
    2018-06-07 06:05:58.01 & 1.75 & 0.587--0.898\\
    2018-06-08 05:17:38.75 & 2.85 & 0.726--0.234\\
    2018-06-17 05:05:12.14 & 1.92 & 0.232--0.573\\
    2019-03-01 08:59:55.17 & 0.1  & 0.551--0.567\\
    2019-07-04 04:04:07.24 & 0.97 & 0.018--0.190\\
    2019-07-05 02:38:01.94 & 2.10 & 0.045--0.419\\
    2019-07-05 23:29:25.96 & 1.56 & 0.766--0.047\\
    2019-07-06 03:18:11.26 & 1.24 & 0.447--0.669\\
    \hline
    \multicolumn{3}{c|}{X-SHOOTER+VLT (Spectroscopy)}\\
    \hline
    2019-06-03 01:14:26.75 & 0.01 & Arclamp\\
    2019-06-03 01:29:46.01 & 1.01 & 0.80--0.97\\
    2019-06-03 02:46:29.93 & 0.01 & Arclamp\\
    2019-06-03 02:50:12.43 & 0.67 & 0.01--0.10\\
    2019-06-03 03:29:23.58 & 0.01 & Arclamp\\
    2019-06-03 05:07:38.08 & 0.01 & Arclamp\\
    2019-06-03 05:17:15.19 & 1.90 & 0.45--0.8\\
    2019-06-03 07:20:15.08 & 0.01  & Arclamp\\
    \hline
    \end{tabular}
\end{table}

The flux from \psr\ was calibrated relative to nearby stars which have measured Pan-STARRS1 (PS1; \citealt{PS1_main}; \citealt{PS1_photo}; \citealt{PS1_data}) magnitudes. The PS1 magnitudes were converted to SDSS magnitudes following the procedure given in \cite{2016ApJ...822...66F}.

Since there are no \textit{u\textprime}\ magnitudes in PS1, the $u_{\rm s}$ zero point of the data taken on 2019-07-04 (the night during which observing conditions were most stable) was found using observations of the standard star LP 617--36 during the same observing run. This zero point was then used to calculate the $u_{\rm s}$ mags of the reference stars in the field of \psr, and these calculated $u_{\rm s}$ reference magnitudes were then used to calibrate the $u_{\rm s}$ data of \psr\ from each of the other nights of data.

\begin{figure*}
    \centering
    \includegraphics[width=\textwidth]{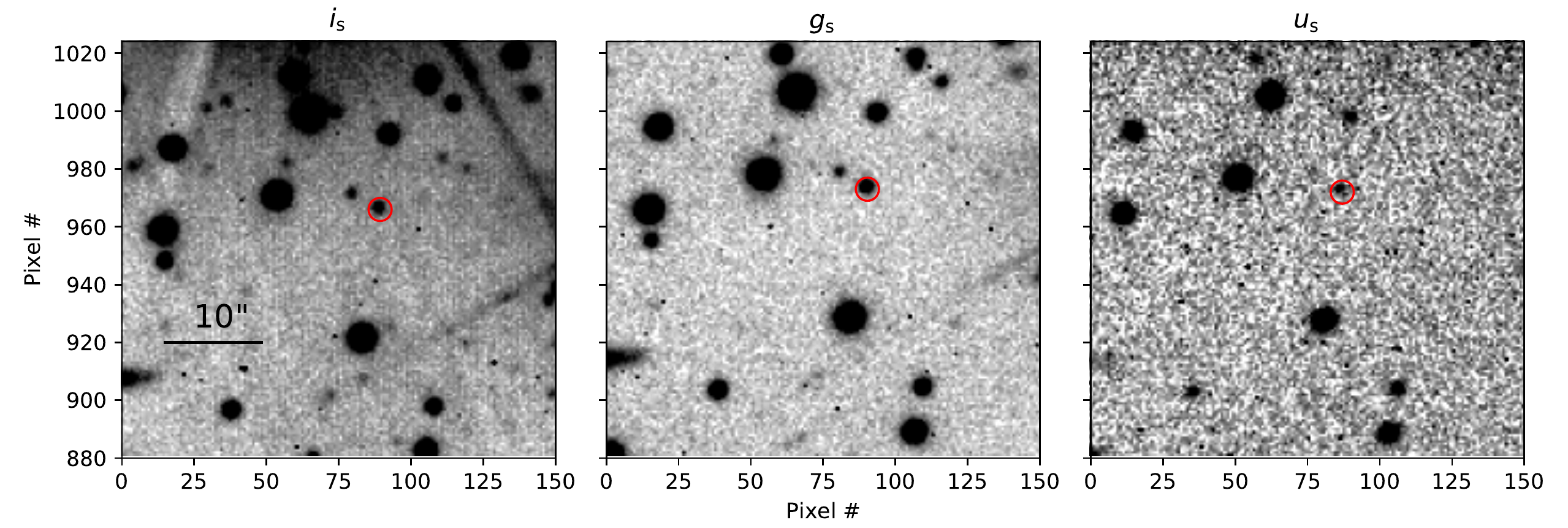}
    \caption{$i_{\rm s}$, $g_{\rm s}$, and $u_{\rm s}$ frames around the position of \psr, which is marked with a red circle. There is a bright star off the top edge of the $i_{\rm s}$ CCD, the artifacts of which prevent accurate band calibration of the $i_{\rm s}$ data from night to night.}
    \label{fig:FoV}
\end{figure*}

\begin{figure}
    \centering
    \includegraphics[width=\columnwidth]{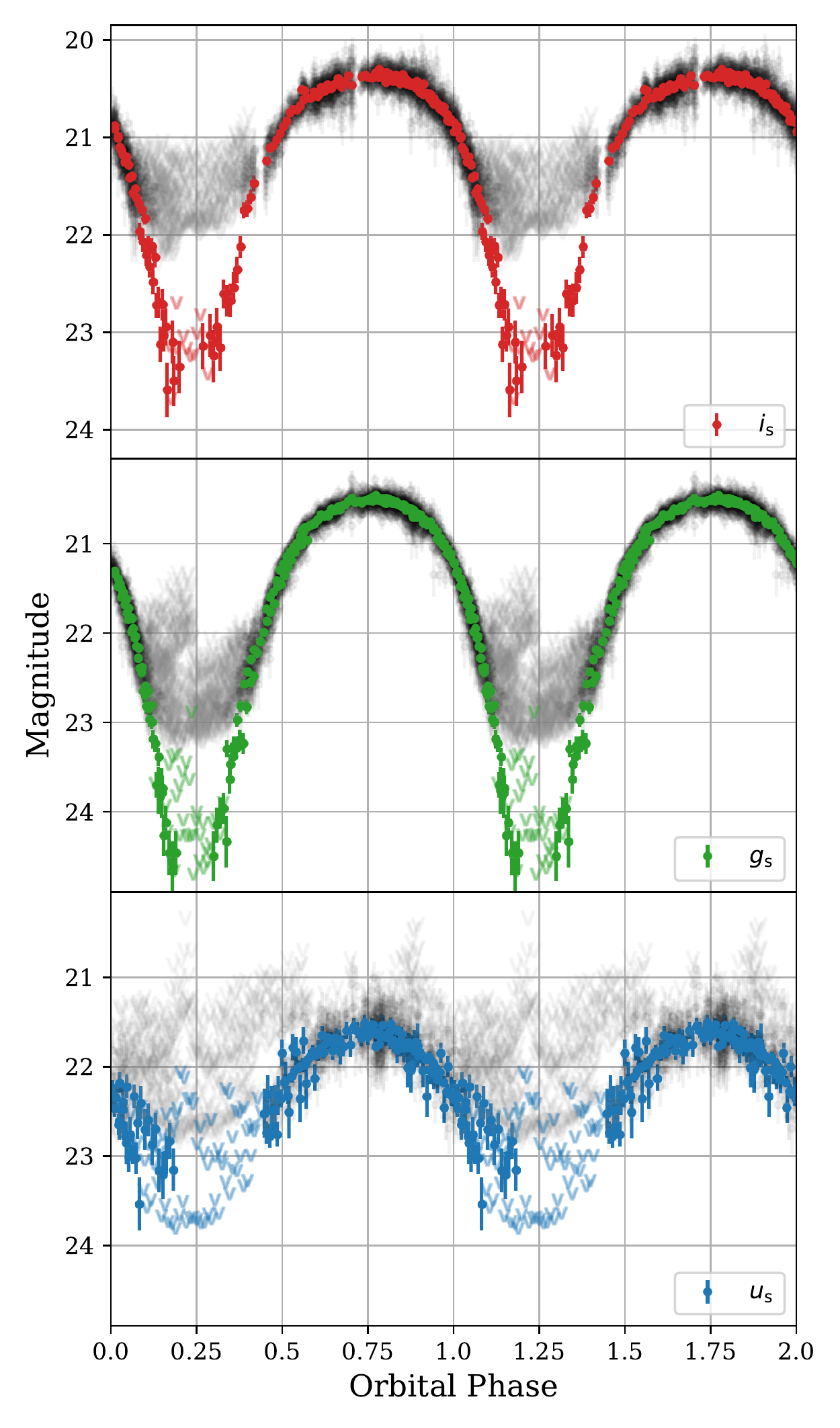}
    \caption{Calibrated light curve of the companion of \psr\ in each of the observed optical bands. Arrows denote upper limits. The grey data points in each plot represent the individual flux measurements from each science frame. The colour points show the data median combined over 200s, allowing us to combine upper limits to reach fainter magnitudes. Orbital phase has been repeated for clarity, and calculated using the parameters given in Table~\ref{tab:results}. Error bars represent 1$\sigma$ uncertainties, and are typically too small to see in the brighter data points at this scale.}
    \label{fig:lc_mag}
\end{figure}

The calibrated $g_{\rm s}$ and $u_{\rm s}$ fluxes agree across every night of observation to within $<0.05$ magnitudes, with the exception of data taken on 2018-06-16, when the data had a shift of 0.2 magnitudes relative to the rest of the observations. The cause of this discrepancy is unclear, but may  be related to the diffraction spike from a nearby star, whose angle changes as the rotator angle of ULTRACAM changes over the course of an observation, passing directly over the target at this time. There are systematic offsets between every night of $i_{\rm s}$ which arise due to the effects of a nearby bright star on the $i_{\rm s}$ CCD. Figure~\ref{fig:FoV} shows a cutout of the CCD around the target in all three bands to highlight the contamination of the $i_{\rm s}$ CCD, and Figure~\ref{fig:lc_mag} shows the $u_{\rm s}$, $g_{\rm s}$, and $i_{\rm s}$ light curves in magnitudes after extraction and calibration.

We also include the SDSS-$r$ band light curve of the companion to \psr\ presented in Ray et al. (2021, submitted). This light curve was not used during the modelling process discussed in subsequent sections. This is primarily because while we can verify that the colours of the source were the same during the various ULTRACAM observations due to the multi-filter nature of the instrument, we cannot do so for the single-filter SDSS-r \textprime light curve of Ray et al.  While it was taken reasonably close in time to the ULTRACAM data, the heating effect in spiders is known to vary (see for example \citealt{2016ApJ...816...74B} and \citealt{2020MNRAS.tmp.3423C}), so we deem it safer not to include the Ray et al. r\textprime\ light curve in our fits. Instead, we only use it to check our best fits to the ULTRACAM data can reproduce it.

\subsection{Optical spectroscopy}\label{sec:VLT_obs}
Spectroscopy of \psr\ was obtained on the night beginning 2019-06-02 using the X-SHOOTER instrument \citep{2011A&A...536A.105V} on Unit 2 (Kueyen) of ESO's 8.2m Very Large Telescope (VLT) as part of ESO observing program 0103.D-0878(A). X-SHOOTER has three Echelle spectrographs covering three different wavelength ranges: the Ultraviolet-Blue (UVB; 2930\AA-5560\AA), Visible (VIS; 5250\AA-10010\AA), and Near-infrared (NIR; 9820\AA-23770\AA) regimes. The exposure times of these cameras were 467, 479, and 490 s respectively. Both the UVB and VIS instruments read out modes were set to 100k/1pt/hg/2x2\footnote{This means a readout mode of 100 kpixels per second with high gain and 2x2 binning. To understand this mode fully, see the X-SHOOTER instrument paper \citep{2011A&A...536A.105V}.}to minimise readout time, leading to dead times of 19 s and 25 s respectively. With this set up, the time between the start of sequential exposures was $\sim$9 min. A 0.9\arcsec slit was used for the VIS/NIR arms, while a 1.0\arcsec slit was used for the UVB arm. This setup led to a spectral resolution (velocity resolution) of 0.8 \AA\ (56 km s$^{-1}$) at 4300 \AA\ in the UVB arm, 0.9 \AA\ (34 km s$^{-1}$) at 7850 \AA\ in the VIS arm, and 3 \AA\ (51 km s$^{-1}$) at 17500 \AA\ in the NIR arm.

The observations were split into two main blocks. The first epoch of observations consists of 12 exposures covering orbital phases 0.8 - 0.1 (where orbital phase 0.0 is defined at the pulsar ascending node). At this point, a brief break was taken as the target approached time of minimum light and was too faint for useful spectra to be obtained. The second epoch encompasses 16 spectra covering 0.45 - 0.8 orbital phase when \psr\ was increasing in brightness. Seeing over the observations varied from 0.4-0.9\arcsec. Full details (including the UTC times of observations) are given in Table~\ref{tab:obs_details}, which also gives the times when arc lamps were obtained, leading to small gaps in the phase coverage.

The spectra were reduced employing the public release of the \textsc{esoreflex} official X-SHOOTER pipeline (ver. 2.9.1; \citealt{esoreflex}) with standard input parameters. Because the target trace becomes particularly faint during orbital phases 0.0-0.5 (the pulsar's superior conjunction occurs at orbital phase 0.25), we used manual extraction mode for the localization of the trace on the image, imposing a fixed centre and half-width of $1.5\, {\rm arcsec}$ for the aperture. The spectra were flux-calibrated using the provided flux standard stars. Finally, as the spectra are reduced within the pipeline in the topocentric reference frame, we proceeded to correct each spectrum from the Earth velocity at the observation time, converting them all to the barycentric reference frame.

\begin{figure*}
    \centering
    \includegraphics[width=\textwidth]{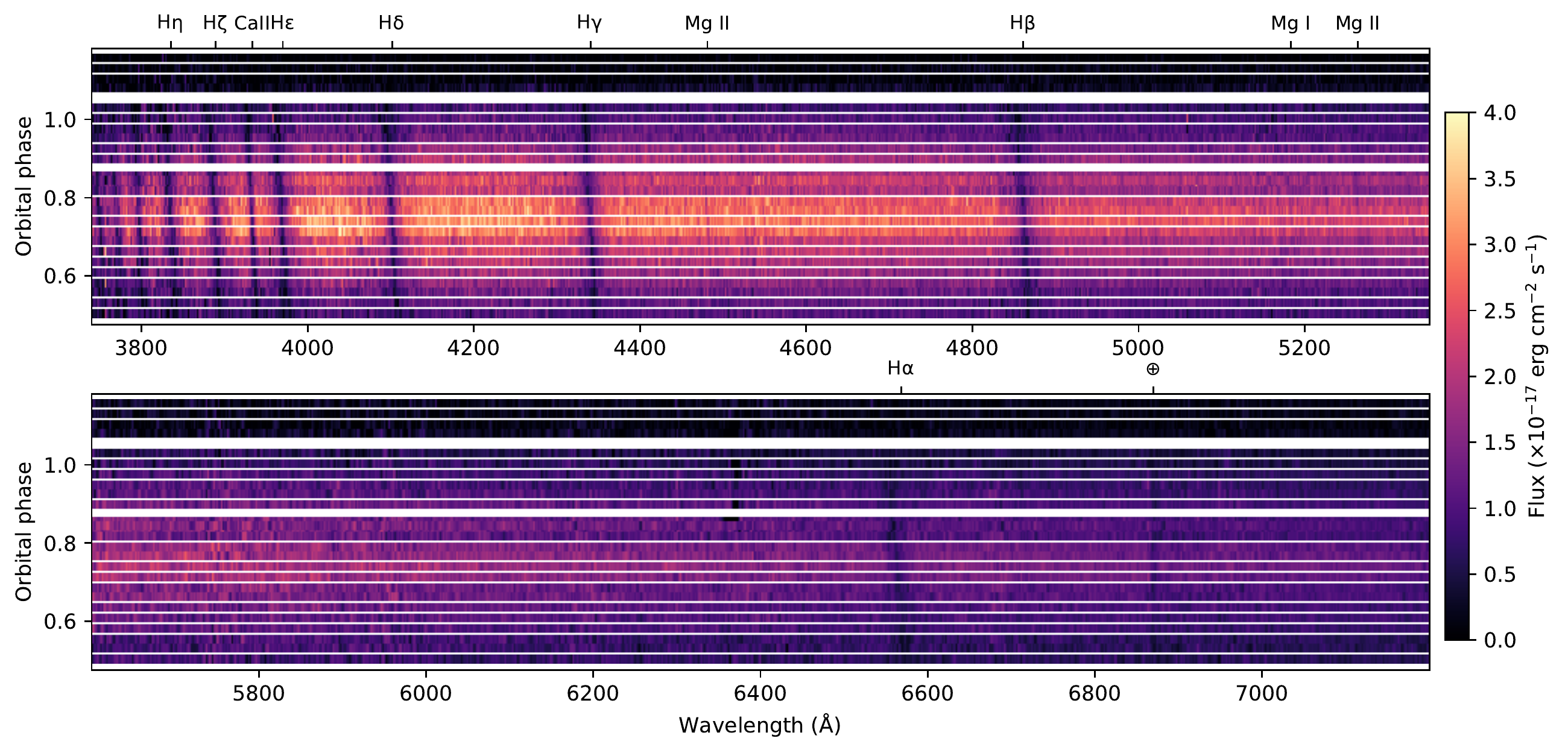}
    \caption{The trailed spectra of \psr\ obtained using the UVB (top) and VIS (bottom) arms of the X-SHOOTER instrument. We have marked the positions of the main transition lines on the top axes of each of the trailed spectrum for clarity. The Hydrogen Balmer series and Ca II at 3933.66\AA\ are clearly visible in absorption. The Ca II line at 3968.47 \AA\ is blended with the H$\epsilon$ line. The spectra convolved with a Gaussian of width $\sigma=3$ pixels for clarity. There are several short gaps in the data which occur between exposures, and two long gaps which occurred when the source was too faint for detection, or when wavelength calibration spectra were being obtained.}
    \label{fig:trailed_spec}
\end{figure*}

Figure~\ref{fig:trailed_spec} shows the UVB and VIS spectra trailed versus orbital phase, where the phase has been calculated using the ephemeris given in Table~\ref{tab:results}. The data taken using the NIR arm are not discussed in this paper, as the source was not detected at these wavelengths at any orbital phase.

\section{Modelling of the Photometry and Spectroscopy}\label{sec:modelling}

\subsection{Priors}
The optical light curves and spectra were modelled using the binary stellar synthesis code \textsc{Icarus} \citep{breton12} along with stellar atmosphere grids computed using \textsc{Atlas9} \citep{2003IAUS..210P.A20C} (see Appendix \ref{sec:spectro_cal} for further details). In the following we have assumed that the companion is tidally locked.

The parameters which were fit for using \textsc{Icarus} were the inclination of the system $i$, the Roche lobe filling factor $f_{\rm RL}$ (measured as the ratio of the stars radius to the L1 point along the line joining both), the base temperature of the star $T_{\rm base}$, the irradiation temperature $T_{\rm irr}$ \footnote{The flux from a given cell in \textsc{Icarus} is computed as $F=F_{\rm base}+F_{\rm irr}$, where $F_{\rm base}=\sigma T_{\rm base}^4$ is the flux due to the base temperature of the star, and $F_{\rm irr}$ is irradiating flux caused by heating of the companion star, and is parameterised as $\sigma T_{\rm irr}^4$. See \citet{breton12} for further information.}, the systemic velocity of the system $\gamma$, the radial velocity amplitude of the companion star $K_2$, the distance to the source $d$, and the optical reddening in the direction of the source $E({\rm g-r})$. We also include parameters to account for a stationary absorption line present in the spectra of \psr\ which we attribute to interstellar absorption at the Ca II doublet, and hyper parameters to account for correlation between neighbouring bins in the spectroscopy as discussed in the appendix.

A prior that is uniform in cos($i$) was applied to the inclination, in line with a distribution in orbital angular momentum direction that is isotropic over the sphere. The distance prior is constructed from combining the expected density of Galactic MSPs along the line of sight towards \psr\ as taken from \cite{2013MNRAS.434.1387L} with the distance estimate of 7.55 kpc inferred from the dispersion measure of the pulsar's radio signal (Ray et al., 2021, submitted; assuming the Galactic electron density model of \citealt{YMW16}). This same procedure was done for the redback system PSR J2039--5617 \citep{2020MNRAS.tmp.3423C}. The prior information for $E({\rm g-r})$ comes from the dust maps of \cite{2019ApJ...887...93G}, which show that $E({\rm g-r})=0.12^{+0.03}_{-0.02}$ for stars between 1 and 8 kpc away in the direction of \psr (1 $\sigma$ error). 

We also allow for band calibration offsets with an amplitude of up to 0.05 magnitudes between each night (except for the previously mentioned data from 2018-06-16, for which we allow up to 0.2 magnitudes), and between our data and model fluxes to account for calibration uncertainties.

\subsection{Procedure}
The photometry and spectroscopy are capable of constraining different sets of the orbital parameters. While a joint fit of both data sets is possible, it can take a significant period of time ($\sim$ weeks to months; see Appendix~\ref{sec:spectro_cal} for specifications of the equipment this was tested on) to converge to a solution and explore the parameter space sufficiently to allow parameter uncertainties to be estimated. The dominant factor in the computing time is calculating the model spectra for each set of parameters, which takes several seconds per likelihood evaluation (for details see Appendix~\ref{sec:spectro_cal}), while several million likelihood evaluations are typically required for parameter estimation.

In contrast, the evaluation of the photometry model is far faster. As such, we first fit the photometry with each of the physical models discussed below, with an uninformative top-hat prior on the radial velocity of the companion star that is 1 inside the bounds ($100<K_2<500$) km s$^{-1}$ and 0 otherwise. This range, in combination with the pulsar's projected semi-major axis and the binary orbital period, equates to a restriction on the neutron star mass of $0.5<M_{\rm NS}<3.5$ M$_{\odot}$. The parameter space was explored to ensure the fit had found the global minimum when considering the photometry data and to estimate parameter errors using \textsc{Multinest} (\citealt{MN1}; \citealt{MN2}; \citealt{MN3}) as implemented in \textsc{Python} through \textsc{Pymultinest} \citep{pymultinest}. \textsc{Multinest} provides the Bayesian evidence, ($Z$), which is used to calculate the Bayes factor and compare models between each other. The Bayes factor is given by
\begin{equation}
B_{1,2} = \frac{Z_1}{Z_2} \,.
\end{equation}
Following the criteria set out in \cite{1939thpr.book.....J}, a ratio in evidence of $<1$ suggests model 2 is preferred over model 1, a ratio of 1-10 is weak evidence, a ratio of 10-150 is moderate, and a ratio $>150$ is strong evidence that model 1 is preferred over model 2. This follows the convention set out by \cite{2009bmc..book.....H}, where a formal definition of  $Z$ may also be found.

In order to incorporate the posterior of the photometry models into the spectroscopy models, the posterior distributions of all the fit parameters from the photometry modelling were then used as priors for the fit parameters when fitting the spectroscopic data. This ensures that when modelling the spectroscopy data, parameters which the spectroscopy fit are insensitive to are sampled in a similar way to photometry fitting. The posterior distributions of the fit parameters from the photometry were modelled using a Gaussian mixture model (GMM) as implemented in \cite{scikit-learn}. In order to select the most appropriate number of components for the GMM, we iterated through 1 to 20 components, and selected n=16 as the most appropriate value as it gave the lowest Bayesian Information Criteria (BIC; \citealt{1978AnSta...6..461S}) score. The spectroscopy data were then fit using \textsc{Multinest} with this GMM prior. We restricted this fitting procedure to the UVB arm of the spectroscopy data to save on computation time, choosing this arm due to the higher S/N and number of stellar features when compared to the VIS arm. The best fit model had a $ \chi ^2$ of 495284.42 for 234000 degrees of freedom. After whitening the residuals according to the procedure described in Section~\ref{sec:spectro_cal}, this improved to 189668.09. After fitting the UVB arm, we computed the model spectra for the VIS arm using the best-fit model parameters to ensure that the model was consistent with the observed data in this arm.

The mass ratio $q=M_{\rm NS}/M_{\rm Comp}$ was computed using the pulsar's measured radial velocity (known from a preliminary $\gamma$-ray pulsation timing solution which is within the uncertainties of the solution as presented in Ray et al. 2021) and the best-fit radial velocity of the secondary from fitting the spectroscopy. Individual masses were then computed using the mass ratio in combination with the measured inclination and orbital period $P_{\rm orb}$ of the system.

\subsection{Models}

Several different models were fit to the data. The first considered was a simple direct heating model, where the inner face of the companion star is heated symmetrically by the neutron star. This model assumes that there is no redistribution of heat around the companion's surface, and that the trailing face (the side of the star opposite the direction of orbital motion) of the companion star is as equally heated as the leading face (the side facing the direction of orbital motion). The symmetric heating model provides a $\chi^2_{\mathrm R}$ = 1.17 (where $\chi^2_{\mathrm R}$ is the $\chi^2$ per degree of freedom for a model) and predicts a pulsar mass of $M_{\rm NS} = 2.9\pm0.1$. The $\chi^2_{\mathrm R}$ is given purely for completeness, and was not used for model selection. Such a high pulsar mass is at odds with even the most exotic neutron star equation of state (see for example \citealt{2013ApJ...773...11H}). The log-evidence of this model ($\log(Z)$) as estimated by \textsc{Multinest} was $-$5243.349.

Considering the high spin down luminosity of this pulsar, it is likely there is a very large temperature difference between the front and back of the star, and that this energy may be diffused over the companion's surface. In effect, this diffusion enables the terminator region between the night and day side of the companion to extend further around the back side of the companion star than in the usual symmetric heating model and is very similar to the situations encountered when modelling the black widow systems PSR B1957+20 \citep{2007MNRAS.379.1117R}, PSR J1311--3430 \citep{2015ApJ...804..115R}, and PSR J1810+1744 \citep{2021arXiv210109822R}. As such, we allowed for diffusion across the surface of the star as discussed in \citet{2020MNRAS.tmp.2692V}, but with an additional component - the strength of diffusion between cells was given a temperature dependence in the form of a power law with index $\Gamma$ (for further information on this extension of the model presented in \citealt{2020MNRAS.tmp.2692V} see \citealt{2021MNRAS.507.2174S}). This model resulted in a $\chi^{2}_{\mathrm R}$ = 1.09, and had substantially stronger evidence than the direct heating model ($\log(Z)$=$-$4877.58 versus $\log(Z)$=$-$5243.349). 

To see if further improvement of the fit to the data could be achieved, we tested an additional source of heating on the companion's surface, as the residuals after subtracting our best fit diffusion model suggested there may be an asymmetry present in the light curve. Such additional sources have been found in many systems previously, and have been ascribed to convection of energy around the surface of the star (\citealt{2020MNRAS.tmp.2692V}; \citealt{2020ApJ...892..101K}), intra-binary shocks \citep{2016ApJ...828....7R}, and cold and hot spots on the surface of the secondary (\citealt{2016ApJ...833L..12V}; \citealt{2020MNRAS.tmp.3423C}). We tested both the convection model with a Gaussian velocity profile, and the hot spot model for \psr. While the slight asymmetry in the residuals was less obvious with these models, both the evidence and $\chi_{R}^2$ of the fit prefer the simple diffusion model over both of these asymmetric models but, most importantly, the asymmetric models gave the same binary parameters for \psr\ as was found using the diffusion model discussed in the previous paragraph. 

\begin{figure*}
\centering
\subfloat{
   \centering
   \includegraphics[width=0.95\textwidth]{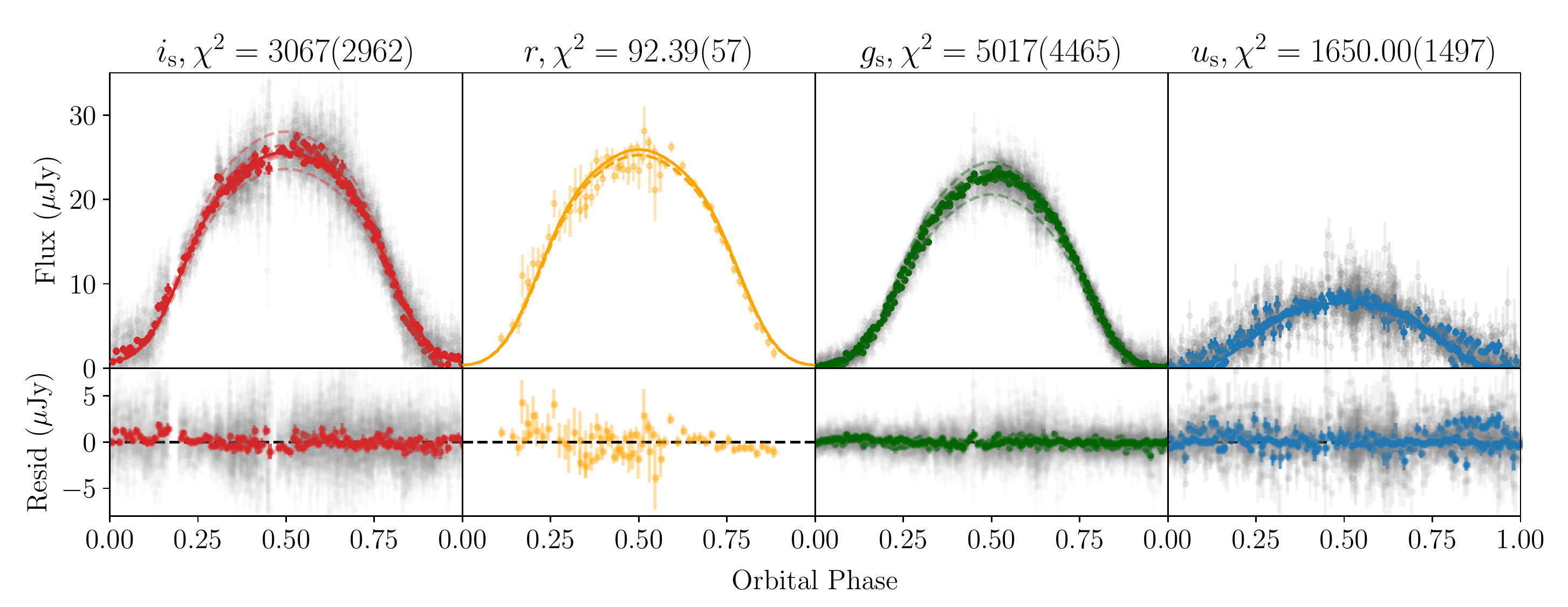}
}

\subfloat{
   \centering
   \includegraphics[width=0.95\textwidth]{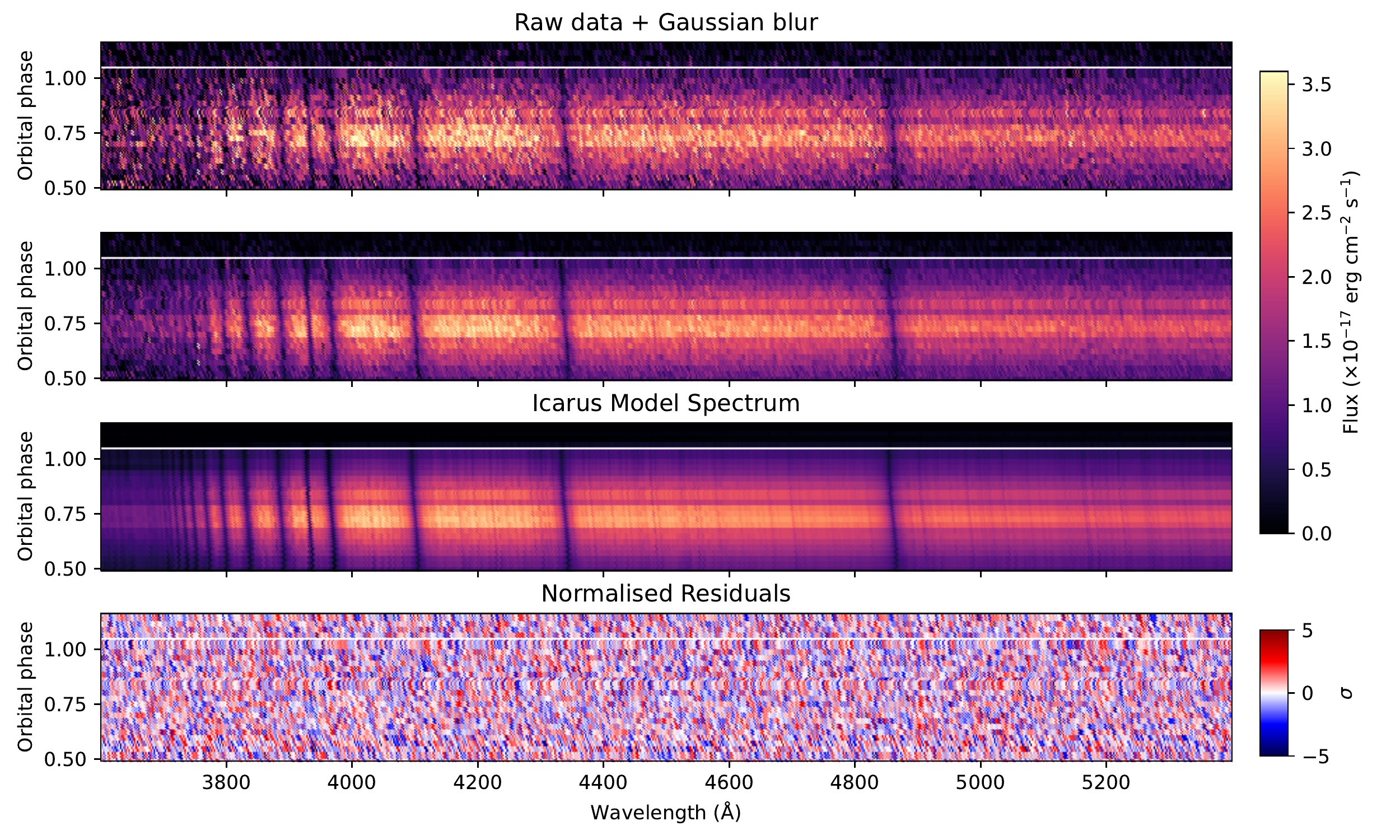}
}
\caption[Two models]{Model fits and residuals to the $u_{\rm s}$, $g_{\rm s}$, \textit{r}, and $i_{\rm s}$ light curves (top) and spectroscopy (bottom) of \psr. The number in brackets beside the $\chi^2$ in each subplot is the degrees of freedom for that band of data. The solid line in each light curve plot shows the theoretical model in that band, while the dashed line (most easily seen in the $i_{\rm s}$ plots) shows the model after accounting for band calibration uncertainties in each night of data. The gray data are the raw data which the model is fit to, while the coloured solid data points show the data after binning each night of data into 100 s exposures, and is done for illustrative purposes only. For the spectroscopy, the top plot shows the raw data and the second plot shows the data convolved with a Gaussian to smooth pixel-to-pixel variations for illustrative purposes only. The third plot shows the model spectrum for each observation after rescaling of the model to match the data, and the bottom plot shows the residuals from subtracting these models from the raw data.}
\label{fig:Model_Results}
\end{figure*}

\begin{figure}
    \centering
    \includegraphics[width=\columnwidth]{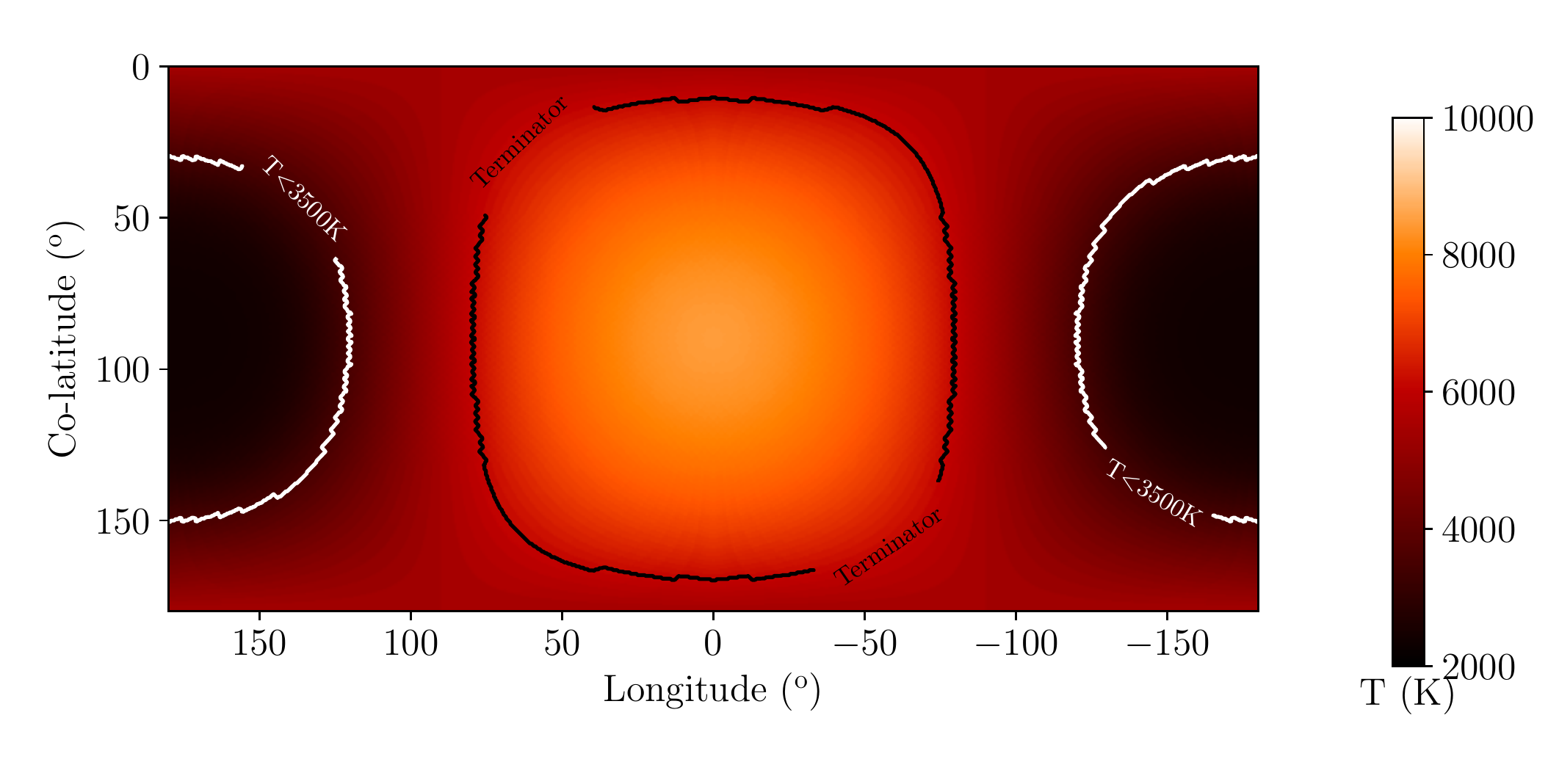}
    \caption{The temperature distribution maps for the companion star in \psr\ for the diffusion model. Here, the co-ordinates follow the convention set out in Icarus, where (0\degree,0\degree) corresponds to the north pole of the companion, (0\degree,90\degree) is the nose of the star and is the closest part of the star to the L1 point, and (180\degree,90\degree) is the back side of the star. The above coordinates are given in (longitude,latitude). The black contour marks the theoretical location of the terminator assuming no heat redistribution, while the white contour encompasses all cells with temperatures below 3500 K, which is the lower limit of the atmosphere grid. The atmosphere grids have been extrapolated below this temperature, and are unreliable. However, as discussed in the text, this area of the stars surface contributes negligible flux to the light curves and spectra.}
    \label{fig:temp_map}
\end{figure}

The corner plot from the \textsc{Multinest} analysis for the diffusion model is given in Appendix \ref{sec:corner_plot}, while the top panel of Figure~\ref{fig:Model_Results} shows the light curves and residuals for the diffusion model. Finally, Figure~\ref{fig:temp_map} shows a temperature map of the surface after the effects of diffusion have been accounted for.

\begin{table*}
    \centering
    \caption{The best posterior parameters from fitting the optical light curves and spectra. The parameter values from photometry acted as prior values for the spectroscopy fitting. Errors are quoted at the 95\%\ level (the errors in brackets are at the 68\%). The base temperature is not the same as the coldest temperature element which occurs on the surface of the star due to diffusion effects. Similarly, the hottest surface element is not equal to a simple combination of the base and irradiating temperatures. As such, we provide the coldest and hottest surface elements in each model after irradiating and diffusion effects have been accounted for. The $\gamma$--ray timing parameters come from Ray et al. 2021 (submitted). The $\chi^2$ value for the spectroscopy fit is the value after whitening the residuals using the Gaussian Process kernel described in Appendix~\ref{sec:spectro_cal}}.
    \begin{tabular}{l c c c c c c}
        \hline
        Parameter               &                                   &                       &                       &               &  \\
                                &  Photometry                       &68\%c.l.               &95\%c.l.               & Spectroscopy  &68\%c.l.           &95\%c.l.\\
        \hline\hline
        $E ({\rm g-r})$                 & 0.15                      &$\pm0.02$              &$\pm0.04$              &               &                   &\\
        $d$ (kpc)                       & 5.1                       &$^{+0.5}_{-0.7}$       &$^{+0.8}_{-1.1}$       &               &                   &\\
        $i$ (\degree)                   & 79                        &$^{+3}_{-2}$           &>75                    &               &                   &\\
        $\gamma$ (km s$^{-1}$)          & -                         &                       &                       &3              &$\pm1$             &$\pm2$\\
        $K_2$ (km s$^{-1}$)             & -                         &                       &                       &397            &$\pm$2             &$\pm$4\\
        $f_{\rm RL}$                    & 0.98                      &$^{+0.01}_{-0.02}$     &$^{+0.02}_{-0.03}$     & 0.95          &$^{+0.02}_{-0.01}$ &$^{+0.04}_{-0.02}$\\
        $T_{\rm base}$ (K)              & 2600                      &<2900                  &<3100               &               &                   &\\
        $T_{\rm irr}$ (K)               & 9400                      &$\pm$140               &$\pm$260               & 9380          &$\pm$40            &$\pm$80\\
        $\kappa$ (W K$^{-1}$ m$^{-2}$)  & 1$\times10^5$             &$\pm0.1\times10^5$     &$^{+0.3}_{-0.2}\times10^5$ &               &                   &\\
        $\Gamma$                        & 6.3                       &$\pm0.4$               &$^{+0.6}_{-0.8}$       &               &                   &\\
        $\lambda_{\rm Abs}$ (\AA)       & -                         &                       &                       & 3933.4        & $\pm0.1$          &$\pm0.2$\\
        $\sigma_{\rm Abs}$ (\AA)        & -                         &                       &                       &0.6            & $^{+0.2}_{-0.1}$  &$^{+0.4}_{-0.2}$\\
        $D_{\rm Abs}$                   & -                         &                       &                       &0.5            &$\pm0.1$           &$\pm0.2$\\
        \hline
        \multicolumn{7}{c|}{Fit Statistics}\\
        \hline
        $\chi^2$                        & 9735.63 (8942)            &                       &                       &189668.09 (234000)              &                   &\\
        $\log(Z)$                       & -4877.58                  &                       &                       &-              &                   &\\
        \hline
        \multicolumn{7}{c|}{Derived Parameters}\\
        \hline
        $q$                             &                           &                       &                       &28.0           &$\pm$0.1           &$\pm$0.3\\
        $M_{\rm NS}$ (M$_{\odot}$)      &                           &                       &                       &1.67           &$^{+0.07}_{-0.05}$ &$^{+0.15}_{-0.09}$\\
        $M_{\rm Comp}$ (M$_{\odot}$)    &                           &                       &                       &0.060          &$^{+0.003}_{-0.002}$&$^{+0.005}_{-0.003}$\\
        $T_{\rm min}$ (K)               &                           &                       &                       &2122           &                   &\\ 
        $T_{\rm max}$ (K)               &                           &                       &                       &8720           &                   &\\
        $\varepsilon$                   &                           &                       &                       &0.32           &$\pm$0.007         &$\pm$0.01\\
        \hline
        \hline
        \multicolumn{7}{c|}{$\gamma$-ray Pulsation Timing Parameters}\\
        \hline   
        $T_{\rm asc}$ (MJD) &             \multicolumn{6}{c|}{57785.539361(1)}\\
        $P_{\rm orb}$ (day) & \multicolumn{6}{c|}{0.2335002682(2)}\\
        $x$ (light second) &\multicolumn{6}{c|}{0.151442(6)}\\ 
        \hline
    \end{tabular}
    \label{tab:results}
\end{table*}

\section{Discussion}\label{sec:Disc}

\subsection{Binary inclination and companion radial velocity}
Modelling of the optical light curve and spectroscopy constrains the inclination of the binary to be $>75$\degree\ at 95\% confidence. Such a high inclination is supported by $\gamma$-ray observations of \psr. The $\gamma$-ray pulse profile shown in Ray et al. (2021, submitted) shows two narrow peaks which are separated by half a rotation of the NS. When compared to the theoretical pulse profiles from \cite{2009ApJ...695.1289W}, this distinct $\gamma$-ray pulse profile suggests the pulsar's spin axis is close to being perpendicular to our line of sight. Coupled with the theory that the spin axis should have been aligned with the orbital axis during the period of accretion which spun up the pulsar, the $\gamma$-ray pulse profile supports the high inclination found through modelling of the optical data.

Traditionally, the radial velocity of the companion star in a binary is estimated by measuring the motion of a set of absorption lines over the orbit. This radial velocity measurement is not necessarily the velocity of the centre of mass of the companion. Rather, it is that of the centre of light for that particular set of absorption lines, and must be multiplied by a correction factor to obtain the radial velocity of the centre of mass when the star is subjected to extreme irradiation, as first done in \cite{1988ApJ...324..411W} and shown in detail for spiders in both \cite{2011ApJ...728...95V} and \cite{Linares2018}. Alternatively, template spectra for a range of spectral types must be cross correlated with the observed spectrum at different times in the orbit.

The method presented here overcomes this limitation by generating the exact spectrum an irradiated star should produce, therefore accounting intrinsically for any change in the centre of light with respect to the centre of mass. This allows us to directly measure the radial velocity of the centre of mass without the need for a correction factor. The modelling presented above constrains the radial velocity of the companion, $K_2$, to be 397$\pm$4 km s$^{-1}$. 

It should also be noted that this technique has advantages over techniques such as Roche tomography, which is capable of reconstructing an image of the surface of the secondary (\citealt{1994A&A...288..773R}; \citealt{2001MNRAS.326...67W}). This is achieved through reconstruction of the observed line profiles, but is limited to doing this on a line-by-line basis, or by modelling the average absorption line profile which is not appropriate when different line species are coming from different parts of the star.

\begin{figure*}
    \centering
    \includegraphics[width=\textwidth]{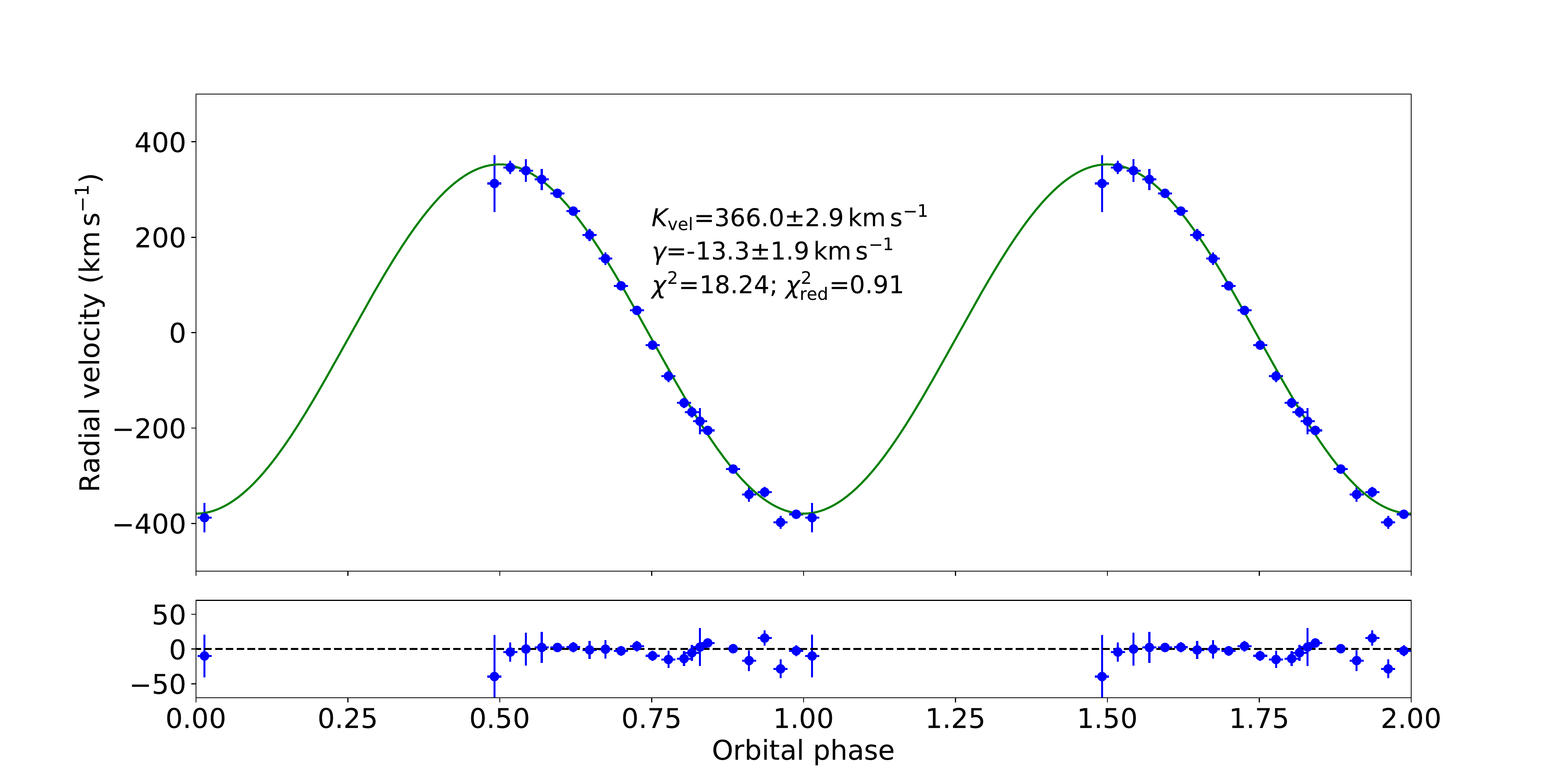}
    \caption{The radial velocity curve for \psr\ obtained using traditional cross-correlation techniques. Here, the radial velocities of the companion at each orbital phase were measured using the spectral ranges of 3750$-$4500 \AA\ and 4700$-$5200 \AA\ to simultaneously cover the Balmer series and Ca II doublet. The bottom panel shows the residuals in km/s. The significant difference between the radial velocity measured using this technique and the value obtained using our new method arises due to the significant heating occurring in this source.}
    \label{fig:RV_curve}
\end{figure*}

When we apply traditional cross-correlation and compare the observed spectra with the normalised synthetic spectrum of a $T_{\rm eff}$=8000 K and $\log(g)$=4.5 star, using the spectral ranges of 3750$-$4500 \AA\ and 4700$-$5200 \AA\ to simultaneously cover the Balmer series and Ca II doublet, we find an amplitude $K_{\rm obs} = 366 \pm 3$ km s$^{-1}$ (see Figure~\ref{fig:RV_curve}). Due to the strong temperature difference between the night and day sides of the star, we expect these species to track the hotter day side of the star, hence the lower value for the radial velocity than the method discussed previously which fits directly for the centre-of-mass velocity. However, care must be taken when interpreting these results, as there may be weak metallic lines within this wavelength range which are highly sensitive to temperature, and indeed the line species we have chosen are themselves highly sensitive to the temperature.

To compare these methods directly, we can predict the minimum ratio of $K_{\rm obs} / K_2 = 1 - f_{\rm eff} R_{\rm nose}/a_2 = 0.85$, following \citet{2011ApJ...728...95V} and where $a_2$ is the distance between the companion star and the centre of mass of the binary. If we assume the nominal mass ratio from Section \ref{sec:mass_ratio}, a Roche-lobe filling companion and all light coming from its innermost point ($f_{\rm eff} = 1$). Our actual value of $366/397=0.92$ falls within this range and implies an effective centre-of-light midway between the innermost point of the companion and its centre-of-mass.

We also attempted a classical approach to determine the rotational velocity of the companion star ($v\sin(i)$) by employing the aforementioned template. We first degraded the synthetic template to the velocity resolution of X-SHOOTER on the VLT under the observational configuration shown in Section~\ref{sec:VLT_obs}. Then, a rotational profile with a linear limb darkening law as defined in \cite{Gray1992} was applied to simulate the rotational broadening for the companion. We compared this template with two spectra: 1) the spectrum obtained at the brightest epoch, and 2) the weighted average of the full normalised set of spectra after correcting them from the  companion radial velocity derived above. In case 1, we obtained best fit results of $v\sin(i)=47^{+10}_{-16}$ km s$^{-1}$ and in case 2 we find $v\sin(i)=64^{+6}_{-7}$ km s$^{-1}$. Case 1 uses only the brightest spectra, which allows us to minimise the effect of the spectral type change over the orbital phase. The better precision of case 2 arises from the higher signal to noise ratio of the averaged spectra, but the effects of the changing spectral type are completely neglected. Both of these results are consistent with the spectral broadening derived from our best fit parameters, which assumed corotation ($v\sin(i)=62$ km s$^{-1}$, \citealt{1988ApJ...324..411W}), but also show the limitations of the classic approach when applied to irradiated spiders.

\subsection{Mass ratio, pulsar mass, and companion mass}\label{sec:mass_ratio}
The measurement of the companion's radial velocity can be combined with the pulsar's semi-major axis of 0.151442 light-seconds to give a mass ratio for \psr\ of $q=M_{\rm NS}/M_{\rm Comp}=28.0\pm0.3$.

\begin{figure}
    \centering
    \includegraphics[width=\columnwidth]{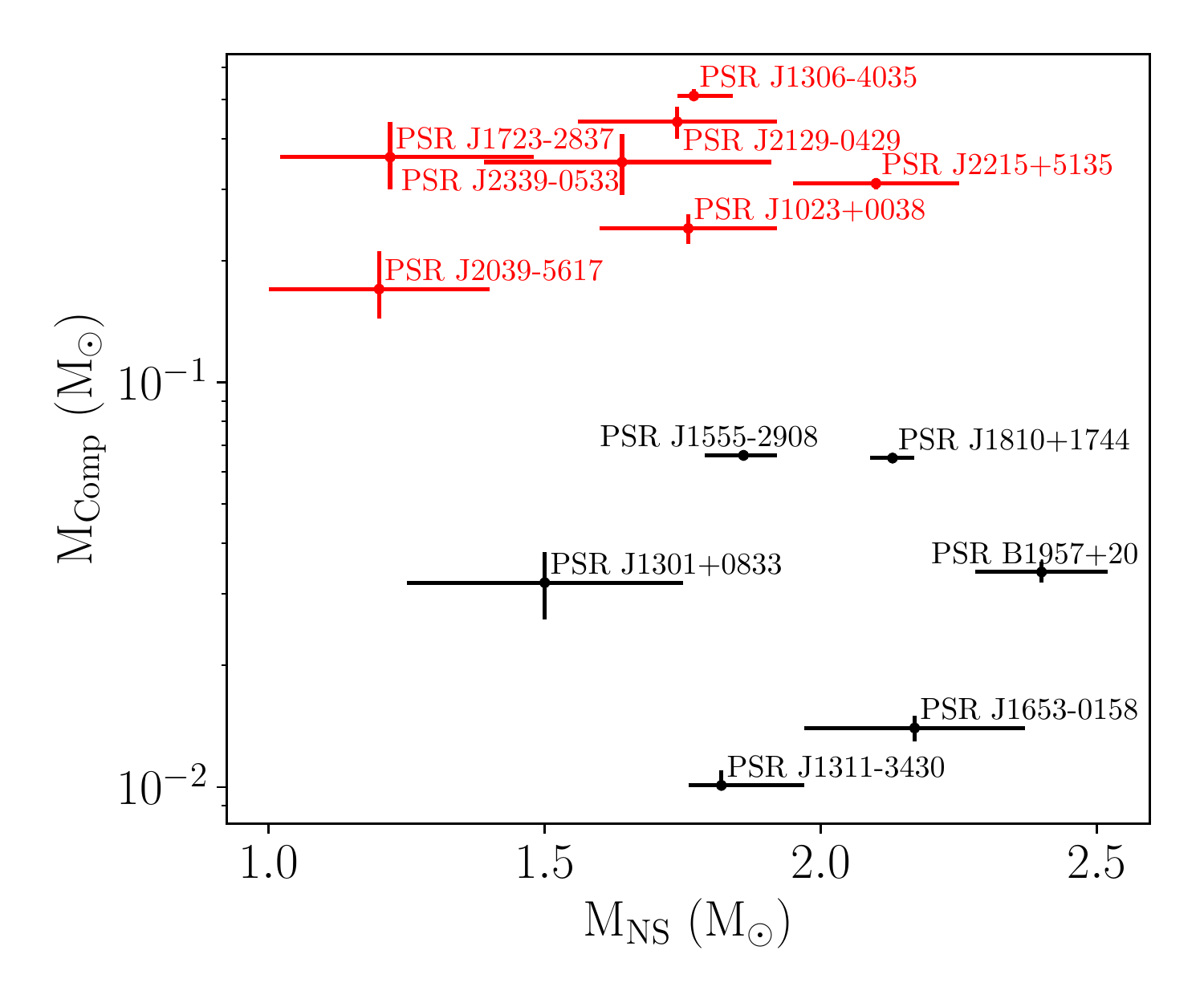}
    \caption{The companion mass versus neutron star mass for all redbacks (red points) and black widows (black points) for which a reliable mass measurement exists. Data for redbacks have been taken from \citet{Strader2019} and references within, while the data for black widows comes from \citet{Linares2019} and references within, with the addition of PSR J1653-0158 from \citet{2020ApJ...902L..46N}, PSR J1810+1744 from \citet{2021arXiv210109822R}, and PSR J2039-5617 from \citet{2020MNRAS.tmp.3423C}.}
    \label{fig:mass_mass}
\end{figure}

Combining this mass ratio with the measured inclination leads to an estimated neutron star mass of $M_{\rm NS} = 1.67^{+0.15}_{-0.09}$ M$_{\odot}$ and a companion mass of $ M_{\rm Comp} = 0.060\pm0.005$ M$_{\odot}$ at the 95\% confidence level. Figure~\ref{fig:mass_mass} shows the component masses for all known black widows and redbacks for which a reliable mass measurement exists.

The mass of the pulsar in \psr\ is typical of those found in other neutron star binaries, however the companion star's mass is one of the heaviest recorded for any black widow system.

An outstanding question surrounding spider systems is whether redbacks and black widows are two distinct populations, and where the boundary between the populations may lie. In previous studies, the boundary has been suggested to lie between 0.05 and 0.1 M$_{\odot}$ \citep{2013ApJ...775...27C}. The factor which dictates whether a system evolves into either a redback or a black widow has been ascribed to the efficiency of the mechanism behind the evaporation of the companion, or perhaps the magnetic field strength of the pulsar \citep{2020MNRAS.495.3656G}. The discovery of heavy companions in both \psr\ and in PSR J1810+1744 suggest the mass gap between the two populations is not as wide as previously thought, and further theoretical modelling of the populations is required to establish whether both populations are independent or related.

\subsection{Companion parameters and heating efficiency}
Both the photometry and spectroscopy suggest the companion star in \psr\ is close to filling its Roche lobe. Modelling of the photometry constrains the filling factor through the strength of the ellipsoidal modulation required to fit the optical light curve, and suggests a filling factor of 0.98$^{+0.02}_{-0.03}$. Additionally, the spectroscopy constrains this parameter through the width of the absorption features in the spectrum, and gives a filling factor of 0.95$^{+0.04}_{-0.02}$. However, given that this filling factor implies a $v\sin(i)$ of 62 km s$^{-1}$ which is comparable to the velocity resolution of 56 km s$^{-1}$ in UVB arm, this suggests the spectroscopy measurement is being dominated by the photometry prior. This filling factor is the ratio of the distance from the star's centre to its nose, measured along the axis joining the companion and the pulsar, to the the distance from the star's centre to the L1 point. The volume averaged filling factor is much higher, with a value $>0.99$.

The base temperature of the companion is less than 3000 K. Such a low temperature is in line with the base temperature of other black widows, but also introduces a caveat to the modelling presented here. The atmospheric grids used in this work to model the photometry and spectroscopy only extend down to 3500 K, as at temperatures lower than this, molecular features begin to dominate the spectrum, and are not treated in the model atmosphere code \textsc{Atlas9} which we used. Fortunately, the parts of the star with temperatures lower than this threshold only occupy a small fraction of the stellar surface (as seen in Figure~\ref{fig:temp_map}). For the best fitting model parameters, we find that the integrated flux from all areas of the surface which have a temperature $T<3500$ K contributes less than 1 $\micro$Jy to the $i_{\rm s}$ light curve, making their contribution negligible to both the model light curves (aside from a narrow range of orbital phase around the minimum) and the spectra.

The inner face of the companion reaches a temperature of 8700 K, consistent with an early A type star. The observed spectra are well fit by a synthetic spectrum which has a metallicity of Z=0. Modelling of the spectra constrain the maximum temperature to be 8720$\pm$80 K. This temperature is likely to be dependant on the underlying model atmospheres which are used to construct the spectroscopy grid. While assessing this dependence is difficult, stellar atmospheres for high temperatures tend to be relatively reliable. With only a grid made using the \textsc{Atlas9} suite of programs available to us to generate the specific intensities which are in turn used to create spectroscopy grid, we are unable to better quantify how strong this dependency may be. As a test, the photometry was modelled using an atmosphere grid constructed using the \textsc{Phoenix} stellar atmosphere code \citep{2013A&A...553A...6H} which extends down to 2,300 K. This grid is not of specific intensities, but rather of integrated spectra, making them unsuitable for spectroscopic modelling. The posterior distributions of all parameters from the \textsc{Multinest} analysis of the photometry using this grid were identical to those found using the \textsc{Atlas9} grid.

There is excellent agreement between the observed spectra and model spectra, with no evidence for any emission lines of any atomic species. This suggests that the heating of the companion's surface is not occurring high in the companion's atmosphere. If it were, then we would expect a thin layer of ionised material to exist in the chromosphere of the star, leading to emission lines in the spectrum like the weak emission features seen in PSR B1957+20 \citep{2011ApJ...728...95V}. Instead, the heating must be occurring deep enough such that the mean free path of any photons emitted due to excitation by the heating source is less than the heating depth, meaning there is no temperature inversion in the atmosphere.

Table~\ref{tab:results} also lists the derived value for the heating efficiency of the system, $\varepsilon=L_{\rm irr}/\dot{E}=0.32\pm0.01$. This is slightly higher than the value of the 0.2 often found in spider systems  \citep{2013ApJ...769..108B}. Higher efficiencies are not unheard of, with the notable example of PSR J2215+5135 for which efficiency has been shown to be highly model dependent ranging from $\sim 0.5$ to $\sim 1.0$ (\citealt{2020ApJ...892..101K}; \citealt{2020MNRAS.tmp.2692V}). There are many reasons why the efficiency can vary so dramatically between systems, but also between models. The heating efficiency depends on both the pulsar wind power effectively intercepted by the cross-section of the companion star, and then the amount of energy that reaches below the photosphere. The first point concerns the denominator $\dot E$, and is biased by the fact that the pulsar wind is actually anisotropic and $\dot E$ should depend on direction. The second point concerns the numerator of $\epsilon$ which depends on the penetration depth of the irradiating high-energy particles \citep[e.g.][]{2020MNRAS.492.1579Z}, and thereby on the composition of the pulsar wind in terms of particle species and their energy spectra, as well as on the structure of the companion's atmosphere above the photosphere. 

\subsection{Source Distance}
Modelling of the photometry suggests the source is at a distance of 5.1$^{+0.8}_{-1.1}$ kpc, with the distance being strongly correlated with the radial velocity amplitude of the companion (see Figure~\ref{fig:corner_plot}). This correlation arises as the photometery constrains the base temperature, irradiation temperatures, and filling factor of the companion. As such, if the distance to the binary is increased, then in order to have the model flux match the observed flux, the physical size of the companion must be increased. Given that the Roche lobe filling factor is close to 1, Icarus can only achieve this by increasing the binary separation, which in turn alters $K_2$.

When combined with the radial velocity of $397\pm4$ km s$^{-1}$ obtained from modelling the spectroscopy, the distance is refined to 5.1$\pm0.2$ kpc. This lies firmly between the 2.6 and 7.55 kpc distance inferred to the source from the radio dispersion measure using to the NE2001 \citep{NE2001} and YMW16 \citep{YMW16} Galactic electron density models, respectively. We note that the 7.55 kpc distance was used in the distance prior, but was assigned a large uncertainty due to the errors associated with distances inferred using the dispersion measure.

The best-fit model prefers a reddening of $E(g-r)=0.15\pm0.04$ which is slightly higher than, but consistent with, the prior range of $E({\rm g-r})=0.12^{+0.03}_{-0.02}$ obtained from \cite{2019ApJ...887...93G}.

\section{Conclusions}\label{sec:Conc}

In this paper, we have presented detailed multi-band optical light curves and spectra of the black widow system \psr, and modelled these data using the binary synthesis program \textsc{Icarus}.

Modelling of the multi-band light curves of \psr\ constrains the binary inclination to be $>75$\degree. The model which best describes the observed data contains a companion in which a significant amount of the pulsar's energy that is absorbed by the companion is redistributed over its surface. There is a large temperature difference of $>6000$ K between the day side and night side of the companion, in line with the very high spin down energy of the pulsar. The companion also seems to be very close to filling its Roche lobe, with the Roche lobe filling factor constrained to be $>0.95$.

Through the use of a new technique for measuring the radial velocity of a star in a binary system, the companion's radial velocity amplitude is found to be 397$\pm$4 km s$^{-1}$. This technique works by generating a synthetic spectrum of the star which accounts for the large temperature difference which exists across the surface, the distortion of the companion's surface due to it filling its Roche lobe, and the viewing angle of the binary. 

The modelling of both the light curve and spectra allows for an accurate measurement of the binary's component masses. The mass of the pulsar, $M_{\rm NS}=1.67^{+0.15}_{-0.09}$ M$_{\odot}$ at the 95\% confidence level, is standard for a pulsar in a spider system \citep{Linares2019}. The mass of the companion in \psr, $M_{\rm Comp}=0.060^{+0.005}_{-0.003}$ M$_{\odot}$, means that it joins PSR J1810+1744 as one of the heaviest black widow companions known to date \citep{2021arXiv210109822R}. Even with these masses, there still appears to be a gap between the companion masses in redbacks and black widows . Further theoretical studies using using binary evolution models will be required to decide if there truly is a mass gap that differentiates black widows from redbacks and, if so, where this mass gap exists.

Finally, the spectroscopic modelling presented here allows for a measurement of the radial velocity amplitude which is fully consistent with the temperature distribution on the companion star's surface. Application of this technique to the published spectra of known black widows and redbacks will be performed in the coming months to see how the published values for their radial velocity amplitudes compares with the values obtained using this method. This will pave the way for neutron star mass measurements in spider systems which are more precise, and hopefully more accurate.

\section*{Acknowledgements}

M.R.K., R.P.B., C.J.C., D.M-S., G.V., and J.S. acknowledge support from the ERC under the European Union's Horizon 2020 research and innovation programme (grant agreement No.715051; Spiders). The proposal to obtain spectroscopy, data acquisition, reduction, and initial manuscript preparation were performed when M.R.K. was a Newton International Fellow funded by the Royal Society (Grant number NF171019). M.R.K also acknowledges support from the Irish Research Council in the form of a Government of Ireland Postdoctoral Fellowship (GOIPD/2021/670: Invisible Monsters) which commenced during the late stages of manuscript preparation. V.S.D., T.R.M., and ULTRACAM acknowledge the support of the STFC. D.M.S. also acknowledges the Fondo Europeo de Desarrollo Regional (FEDER) and the Canary Islands government for the financial support received in the form of a grant with number PROID2020010104.

Based on observations collected at the European Southern Observatory, Chile under ESO programme 0103.D-0878(A). Portions of this work performed at NRL were supported by NASA.

The Pan-STARRS1 Surveys (PS1) and the PS1 public science archive have been made possible through contributions by the Institute for Astronomy, the University of Hawaii, the Pan-STARRS Project Office, the Max-Planck Society and its participating institutes, the Max Planck Institute for Astronomy, Heidelberg and the Max Planck Institute for Extraterrestrial Physics, Garching, The Johns Hopkins University, Durham University, the University of Edinburgh, the Queen's University Belfast, the Harvard-Smithsonian Center for Astrophysics, the Las Cumbres Observatory Global Telescope Network Incorporated, the National Central University of Taiwan, the Space Telescope Science Institute, the National Aeronautics and Space Administration under Grant No. NNX08AR22G issued through the Planetary Science Division of the NASA Science Mission Directorate, the National Science Foundation Grant No. AST-1238877, the University of Maryland, Eotvos Lorand University (ELTE), the Los Alamos National Laboratory, and the Gordon and Betty Moore Foundation.

This work made use of \textsc{Astropy} \citep{astropy:2013,astropy:2018}.

\section*{Data availability}
The raw spectra of PSR J1555-2908 are available through the ESO archive, while the raw ULTRACAM images and associated calibration frames may be obtained by contacting M. R. Kennedy or the ULTRACAM team (V. S. Dhillon or T. R. Marsh). The reduced spectra and reduced light curves are located at a permanent Zenodo repository: \url{https://zenodo.org/record/5653061}.




\bibliographystyle{mnras}
\bibliography{j1555} 



\appendix

\section{Spectroscopic Modelling}\label{sec:spectro_cal}

A paper on generating high resolution synthetic spectra of binary stars with \textsc{Icarus} is in preparation (Breton et al., in prep), and a detailed discussion is beyond the scope of this paper. However, in this section we aim to give a brief overview of the assumptions surrounding the modelling of the optical spectra in this paper.

Before computation of a synthetic spectrum, a grid of specific intensities covering the required range in temperature, surface gravity ($\log_{10}(g)$), cosine of the angle between the normal to the surface and the line to the observer ($\mu$), and observed wavelength is required. The specific intensity grid used in this paper was generated using the \textsc{Atlas9} suite of programs \citep{2003IAUS..210P.A20C} and the available model atmospheres \footnote{\url{http://wwwuser.oats.inaf.it/castelli/grids.html}}. The specific intensities were computed with a resolution ($R=\Delta\lambda/\lambda$) of 300000 from 2500 \AA\ up to 25,000 \AA\, and cover a temperature range in kelvin of $3500<T<13000$ in steps of 250 K, and a surface gravity range of $2.5<\log(g)<5.0$ in steps of 0.5, with solar metallicity assumed. These grids were then degraded to match the resolution of X-SHOOTER on the VLT by convolving the spectra with a truncated Gaussian to represent spectral broadening by the seeing, truncated by the slit. Finally, to compute the model spectrum, the emergent spectrum from each surface element on the tessellated surface of the companion is computed using this grid.

We also account for the effects of smearing of the spectrum due to the orbital motion of the companion star. For a spectra taken at orbital phase $\phi$ and lasting a duration of $d\phi$, the maximum velocity shift between the start and end of the exposure is computed. This velocity is then compared to the velocity resolution of the data. If the expected velocity change over the exposure time is larger than the velocity resolution of the data, then 5 synthetic spectra are computed spanning $\phi-d\phi/2$ up to $\phi+d\phi/2$. The model spectrum is then computed by integrating these 5 spectra.

A single likelihood evaluation requires the flux from 3,072 surface elements to be extracted from the atmosphere models and a 4D interpolation performed at 12,854 wavelengths for each of the 26 spectra, for ~16 billion memory look-ups per likelihood evaluation. To achieve reasonable computation times, we store the degraded atmosphere models on an NVIDIA TITAN RTX graphics card, whose 24GB of memory is sufficient to store the 11GB set of atmosphere models, and have implemented the atmosphere model retrieval and interpolation using the CUDA framework. This is called by \textsc{Icarus} using the \textsc{pycuda} module \citep{kloeckner_pycuda_2012}. With this setup, a full likelihood evaluation takes $\sim30$s.

\begin{figure*}
    \centering
    \includegraphics[width=\textwidth]{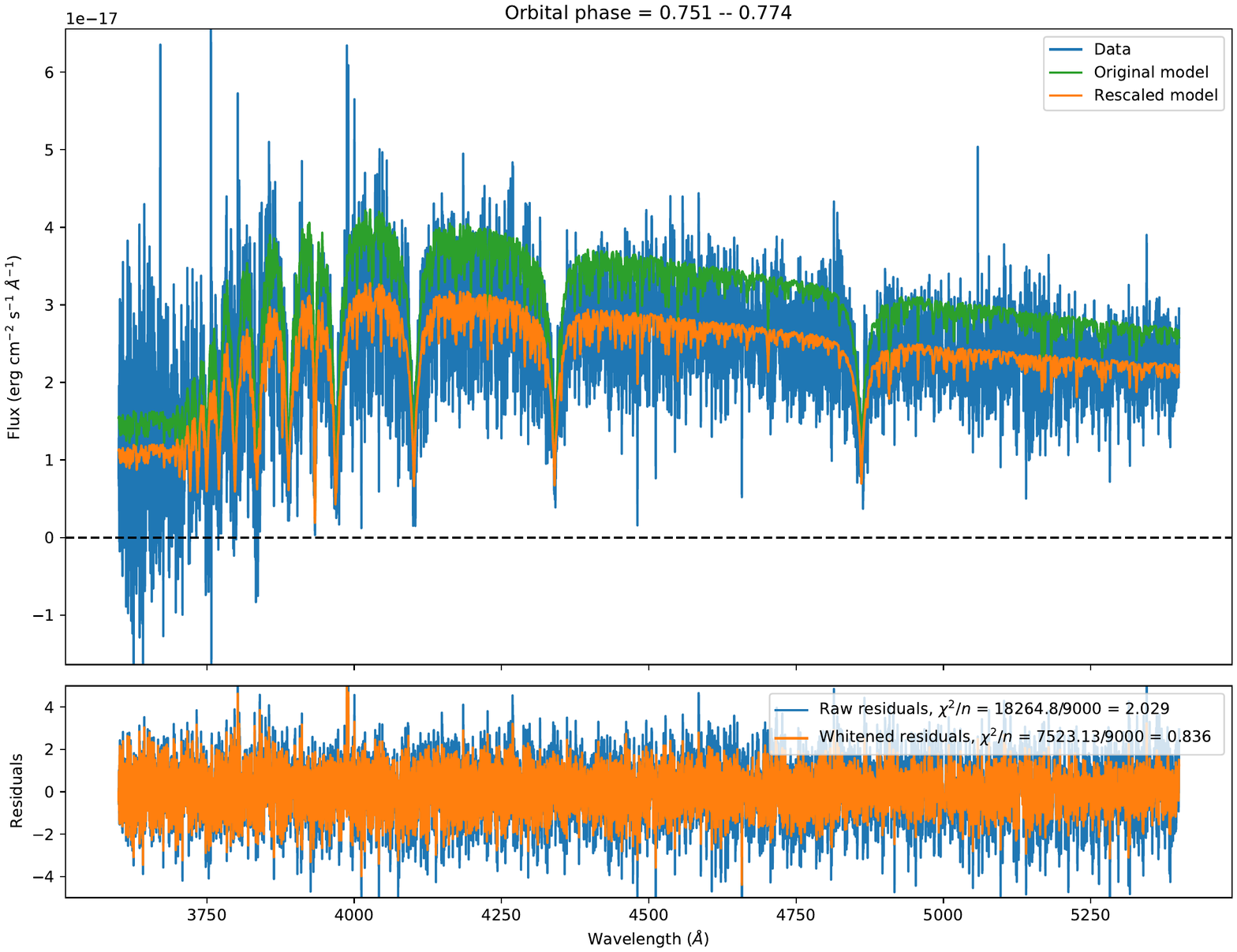}
    \caption{An example of a model spectrum (green) rescaled (orange) to match an observed spectrum (blue). The top panel shows the data and model, while the botom panel shows the residuals before (blue) and after (orange) accounting for the correlated noise using a Gaussian Process.}
    \label{fig:individual_spec}
\end{figure*}

We do not use information from the continuum of optical spectroscopy when calculating the log likelihood of a given model, as flux calibration typically requires fitting a high order polynomial to the observed spectrum of a standard star. The fact that the standard star is observed at different airmass and seeing conditions than the target, combined with potential errors derived from the polynomial fit, compromise the reliability of the target's flux calibration and made them unsuitable to be compared with the continuum of our synthetic grid. Instead, per observed spectrum, we generate a synthetic spectrum which is then rescaled to match the flux calibrated spectrum through fitting of a 4th order polynomial, with no information from the continuum used to constrain any of the orbital parameters. Instead we rely on the information derived from the depth, width, and position of the absorption lines in the spectrum. The parameters that can then be constrained from matching the absorption lines within the observed and synthetic spectra are the parameters related to the temperature of the star, the inclination, the filling factor, the systemic velocity, and the radial velocity of the companion star. Figure~\ref{fig:individual_spec} shows an example of one of the observed spectra along with one of the model spectra after scaling the continuum of the model by the 4th order polynomial.

There is significant correlation between neighbouring pixels in the X-SHOOTER data, which has been ascribed to the co-addition of nodding exposures by the ESO pipeline when producing the individual spectra (a known effect previously reported in other works, see e.g. \citealt{2016A&A...594A..91L}). We model this correlation as a Gaussian Process with a Radial Basis Function kernel of the form
\begin{equation}
    k(x_{\rm i},x_{\rm j}) = h\exp\left(-\frac{d(x_{\rm i},x_{\rm j})^2}{2l^2}\right)
\end{equation}
where $h$ is a hyper parameter controlling the absolute amplitude of the kernel, $k(x_{\rm i},x_{\rm j})$ is the strength of the correlation between datapoints $x_{\rm i}$ and $x_{\rm j}$, $d(x_{\rm i},x_{\rm j})$ is the distance between points $x_{\rm i}$ and $x_{\rm j}$, and $l$ is a hyper parameter controlling the length scale of correlation. Further details on this method will be included in the upcoming paper describing the spectroscopy module. For further reading on Gaussian Process Modelling, see \cite{2006gpml.book.....R}. 

To account for correlations between neighbouring pixels, we ``whiten'' the residuals by transforming with the covariance matrix, $C$,
\begin{equation}
C_{ij} = k(x_{\rm i},x_{\rm j}) + \Delta f_i^2 \delta_{ij},
\end{equation}
where $\Delta f_i$ are the flux uncertainties, and $\delta_{ij}$ is the Kronecker delta function (i.e. the uncertainties are added in quadrature to the diagonal of the covariance matrix.
The likelihood is computed according to Equation (6) of \cite{2015ApJ...812..128C}, which includes a normalisation term that penalises unnecessarily large covariance functions.

\section{Multinest Corner Plots}\label{sec:corner_plot}

\begin{figure*}
    \centering
    \includegraphics[width=\textwidth]{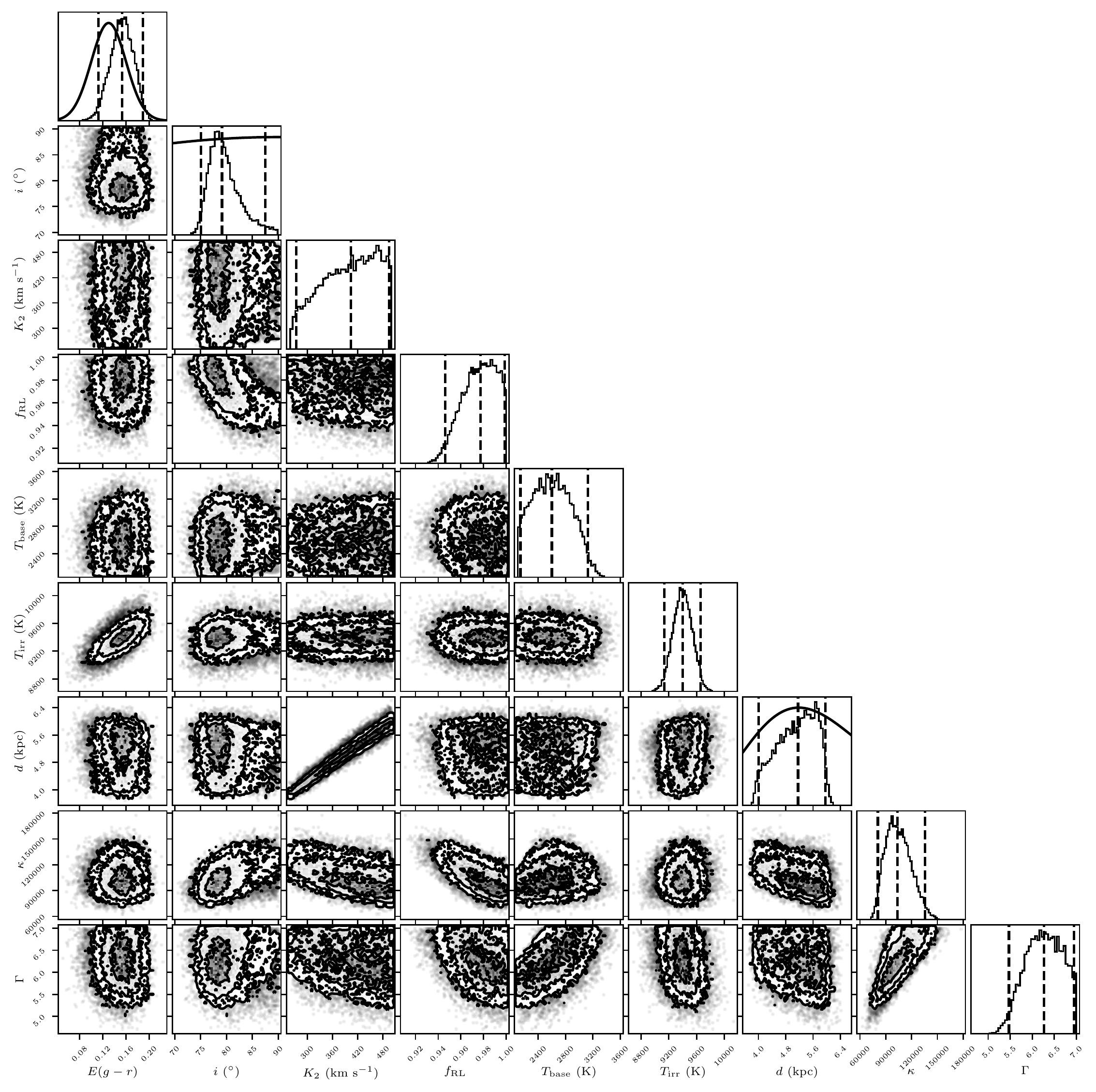}
    \caption{The corner plot generated from the \textsc{Multinest} analysis of the light curve of the companion to \psr. Contours are plotting at the $1\sigma$, $2\sigma$, and $3\sigma$. The priors applied to the specific parameters discussed in the text are plotted as solid curves along the diagonal. The dashed lines mark the 2.5\%, 50\%, and 97.5\% levels of the CDF.}
    \label{fig:corner_plot}
\end{figure*}

\begin{figure*}
    \centering
    \includegraphics[width=\textwidth]{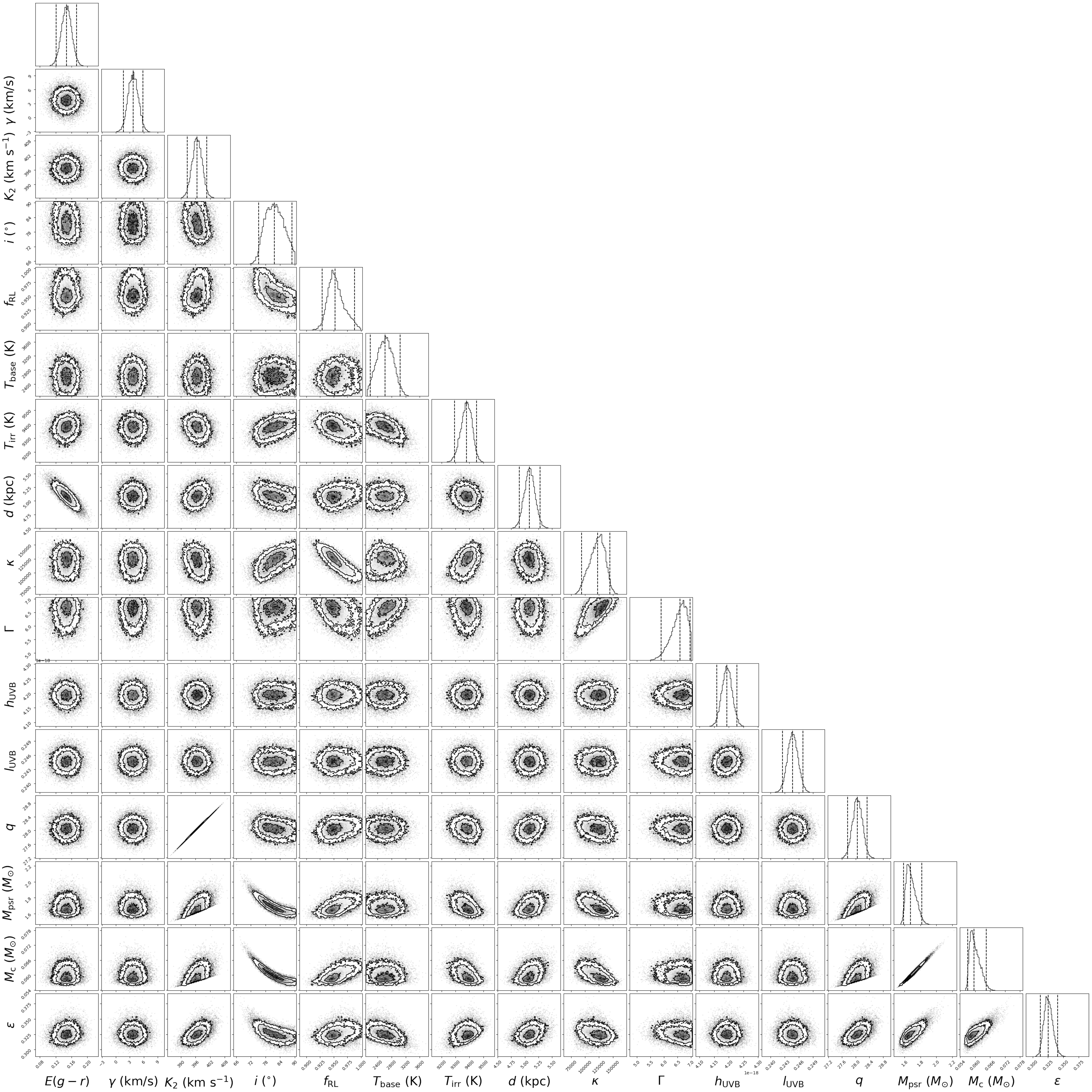}
    \caption{The corner plot generated from the \textsc{Multinest} analysis of the spectroscopy of the companion to \psr. Contours are plotting at the $1\sigma$, $2\sigma$, and $3\sigma$. A prior based on a Gaussian Mixture Model has been applied to all parameters that are in common with the photometry corner plot. The dashed lines mark the 2.5\%, 50\%, and 97.5\% levels of the CDF. $h_{\rm UVB}$ and $l_{\rm UVB}$ are the hyper parameters associated with the Gaussian Process Modelling procedure detailed in Appendix~\ref{sec:spectro_cal}}
    \label{fig:corner_plot_spec}
\end{figure*}

\bsp	
\label{lastpage}
\end{document}